\documentstyle[12pt]{article}
\def\reference{\parskip 0pt\par\noindent\hangindent 0.5 truecm}

\begin{document}
%
%
\title{SUPERNOVA REMNANTS, PULSARS AND THE INTERSTELLAR MEDIUM - SUMMARY OF A WORKSHOP HELD AT U SYDNEY, MARCH 1999}
%


\author{Vikram Dwarkadas $^{1}$ \and
 Lewis Ball $^{1}$ \and
 James Caswell $^{2}$ \and
 Anne Green $^{3}$ \and
 Simon Johnston $^{1}$ \and
 Brian Schmidt $^{4}$ \and
 Mark Wardle $^{1}$ 
} 

\date{}
\maketitle

{\center
$^1$ SRCfTA, School of Physics, A28, Univ of Sydney, NSW 2006\\vikram@physics.usyd.edu.au \\ ball@physics.usyd.edu.au \\ simonj@physics.usyd.edu.au \\ wardle@physics.usyd.edu.au \\[3mm]
$^2$ Australia Telescope National Facility, PO Box 76, Epping NSW 1710 \\jcaswell@atnf.csiro.au\\[3mm]
$^3$ Department of Astrophysics, School of Physics, Univ of Sydney, NSW 2006 \\agreen@physics.usyd.edu.au\\[3mm]
$^4$ RSAA, ANU, Private Bag, Weston Creek P.O., Canberra, ACT 2611\\brian@mso.anu.edu.au \\[3mm]
}

%
\begin{abstract}

	We summarise the proceedings of the SRCfTA workshop on
	``Supernova Remnants, Pulsars and the Interstellar Medium''
	that was held at the University of Sydney on Mar 18 and 19,
	1999.

\end{abstract}


\bigskip

%
%

\section{Introduction}

The study of Supernova Remnants (SNRs) and their interaction with the
surrounding medium has made significant advances in the last decade or
so, thanks in large part to detailed observations of SN 1987A and SN
1993J. The vast amounts of data obtained over several years of study
have considerably improved our understanding of the evolution of young
supernova remnants in general. The coincidence of occurrence of SN
1998bw within the error circle of the gamma-ray burst GRB 980425,
suggesting a relationship between the two objects and new avenues to
advance our understanding of them, has added an exciting new dimension
to our investigation of supernovae (SNe).\\

With a view to discuss the latest results on these and similar topics,
The Special Research Centre for Theoretical Astrophysics at the
University of Sydney organised a workshop on Supernova Remnants,
Pulsars and the Interstellar medium. The two day workshop (March 18-19
1999) brought together more than 65 observers and theorists from all
over Australia (and even a few from overseas), providing a forum for
frank discussion and vigorous interaction. The topic was broadly
interpreted, and the agenda for the meeting was kept open to
accommodate talks that would be interesting to the audience, even if
they did not easily fall into one of the categories. Graduate students
were especially encouraged to attend and present their work.\\

A discussion of supernovae naturally leads one to think of the stellar
remnants that remain after the explosion. In recent years large-scale
surveys have led to a large increase in the number of known
pulsars. Thus pulsars and the nebulae around them formed an important
part of the workshop, with two sessions being devoted to the study of
pulsars and their properties, especially radio pulsars. There were
also interesting reviews presented on contemporary topics such as
Magnetars and Anomalous X-ray Pulsars.\\

Finally the last session was devoted to the study of masers in SNRs, a
field that, after a period of dormancy, is enjoying a great resurgence
nowadays. Intriguing new observations were revealed, with the promise
of more to come.\\

The following summary captures the essence of the science that was
discussed at the workshop. The various sections correspond to the
sessions at the meeting. Further details, and abstracts of talks, are
available at the meeting home page:\\

http://www.physics.usyd.edu.au/$\sim$vikram/snrwkshop/snmain.html \\


\section{Supernova remnants and the Surrounding Medium I}
The first session in this workshop dealt with supernovae (SNe) and their
interaction with surrounding circumstellar material (CSM). In particular,
papers were presented on the diversity of SNe in general, and on some
detailed observations of two very young objects (SN 1987A and SN 1993J)
which now show evidence for interaction between the expanding ejecta and
the surrounding material.\\

It is clear that the evolutionary stage of the progenitor star
determines the kind of SN that occurs. However, it is only rarely that
we have comprehensive data on the progenitor. Typically,
classification is made from the observation of the supernova event and
its aftermath, and a great diversity is seen in these catastrophic
explosions. {\bf Brian Schmidt} (ANU) gave a comprehensive review of
SN classification emphasising this diversity and the fact that many
events do not fit the existing sub-type classifications, based on
studies of light curves and optical spectra (Filippenko 1997).
Because there are so many SNe which are atypical, it may be that the
broader groupings of ``thermonuclear explosions'' (involving
predominantly white dwarfs) and ``core collapse of massive stars''
might lead to better predictions of the progenitor star type. It is
clear that variations in age, mass and metallicity can all affect the
SN light curve and spectrum. In the core collapse scenario, there are
five phases which produce different spectral characteristics. These
phases relate to shock break-out, adiabatic cooling, transfer of
energy and subsequent radioactive heating in the core, and the
eventual transition to the nebula phase. However, it is unclear what
are the primary causes of differences in observed events. Present
models involving variation in the energy of the initial explosion,
mass loss rates and the condition of the CSM do not seem to explain
the observed diversity. In a further twist, it may be that some types
of SNe (for example Type Ib/c) may be linked to the phenomenon of
Gamma-Ray-Bursters (GRBs).\\

The first specific example selected to demonstrate CSM interaction was
SN 1987A, in the Large Magellanic Cloud. {\bf Ray Stathakis} (AAO)
presented results from optical and infrared monitoring with several
instruments mounted on the Anglo Australian Telescope (AAT). Hubble
Space Telescope (HST) images show evidence for the SN ejecta
interacting with the edge of the CSM, from H$\alpha$ and Ly-$\alpha$
observations. At the AAT the source is not well resolved. However,
monitoring of optical CSM lines establishes valuable baseline levels
against which future changes due to increasing interaction may be
measured. Several spectral lines, both in the optical (eg.~ H$\alpha$,
OI) and infrared (eg.~ FeII, Br$\gamma$) regimes are becoming strong
enough to image. It is anticipated this program will continue.\\

Radio observations of SN 1987A, made with the Australia Telescope
Compact Array (ATCA), were presented by {\bf Lister Staveley-Smith}
(ATNF). Evolution of both angular size and flux density is seen. Radio
frequency observations have been an effective way of monitoring the
expanding shock front (Gaensler et al.~1997). Finding a consistent
model to explain the results is more problematic. The data show that
the overall radio luminosity is increasing linearly and that the EW
asymmetry in the brightness of the observed circular ring is also
becoming more pronounced. From the change in image size over several
years, it appears that the expansion velocity has slowed
significantly. The morphology of the images suggests a thin spherical
shell with an EW asymmetry, expanding and now very close to the ring
of CSM. Evidence for the onset of interaction is seen in the HST
H$\alpha$ and Ly-$\alpha$ observations. It is speculated that the
emission is coming from the reverse shock, consistent with the low
value for the expansion velocity. Two possible models which might
explain the observed results both have some unsatisfactory
features. The minimum energy solution is inconsistent with a low shock
velocity and the model invoking a dense HII torus to account for the
slow shock velocity would not predict the symmetric ring observed, nor
the inferred spherical shock. Overall, it seems that SN 1987A was an
atypical Type II SN. It is expected that the shock will heavily impact
the CSM ring in about 2004. High resolution observations at 20 GHz are
planned with the ATCA for the anticipated impressive display.\\

The second object selected to illustrate early interaction with the
CSM is SN 1993J. {\bf Michael Rupen} (NRAO) showed results from VLBI
observations of this young SNR (Bartel et al. 1994; Rupen et
al. 1998), which was the brightest optical SN seen in the northern
hemisphere since 1937. Early observations classified this event as a
core collapse SN (Type IIb) of a massive progenitor star, probably
about 15 M$_{\odot}$. This object has been closely monitored since 30
days after the explosion over several wavelengths in the range 1--20
cm. The SN occurred in M81, a galaxy 3.63 Mpc away. Distance estimates
from the SN observations agree well with the independent estimate from
Cepheid measurements. The object is now seen as a nearly-circular
expanding shell with an asymmetric brightness distribution. There is
some indication that the core may be located off-centre.  However,
there is clear evidence of source evolution and the shell is
noticeably decelerating, even if the most extreme opacity effects are
included. \\

From the review by Schmidt and the specific data on SN 1987A and SN
1993J, it is clear that even for very young remnants, the
peculiarities of the individual SN explosion and the pre-existing CSM
are far stronger influences than any underlying generic
characteristics. This makes it hard to develop global theories and
emphasises the need for continuing searches and subsequent long-term
monitoring of SNe. \\

\section{Supernova Remnants and the Surrounding Medium II}

{\bf Miroslav Filipovic} (U Western Sydney) presented evidence for a
young, nearby SN remnant, RX J0852.0-4622, initially identified by its
X-ray and $\gamma$-ray emission.  He showed that the X-ray image
obtained in the ROSAT all-sky survey shows a disk-like, partially
limb-brightened emission region, which is the typical appearance of a
shell-like SNR. The object's high temperature of $> 3\times 10^7$K
indicates that RXJ0852.0-4622 is a young object which must also be
relatively nearby (because of its large angular diameter of
2$^\circ$).  Comparison with historical SNRs limits the age to about
$\sim$1,500 years and the distance to $<1$ kpc. Miroslav showed that
any doubt of the identification of RX J0852.0-4622 as a SNR should be
erased by the detection of $\gamma$-ray line emission from $^{44}$Ti,
which is produced almost exclusively in supernovae. Using the mean
lifetime of $^{44}$Ti (90.4 yrs), the angular diameter, adopting a
mean expansion velocity of 5000 km/s, and a $^{44}$Ti yield of
$5\times 10 ^{-5}$ M$_\odot$, Iyudin et al. (1998) derived an age of
$\sim 680$ years and a distance of $\sim$ 200 pc, which argues that RX
J0852.0-4622 (GRO J0852-4642) is the closest supernova to Earth to
have occurred during recent human history. However, these observations
are in apparent conflict with historical records. Miroslav also
reported a positive radio-continuum detection at 4.75 GHz (PMN) which
shows similarities to the X-ray emission. Further studies of this SNR
will compare a mosaic radio-continuum survey to observations at other
wavelengths such as the ROSAT and ASCA X-ray images and spectra
(already observed) and UKST H$\alpha$, [SII] and [OIII] plates. \\

{\bf Vikram Dwarkadas} (SRCfTA) presented work he, along with Roger
Chevalier (UVa), is carrying out on SN-circumstellar interaction,
motivated by the presence of a circumstellar bubble surrounding SN
1987A. The evolution of supernova remnants in circumstellar bubbles
depends mainly on a single parameter - the ratio of the mass of the
circumstellar shell to the mass of the ejecta (Franco et
al.~1991). For low values, the supernova remnant, over many doubling
times, eventually `forgets' about the existence of this shell, and the
resulting density profile looks as it would have in the absence of the
shell.  Vikram showed that analytical approximations and numerical
models indicate that the evolution becomes more rapid as this ratio
increases, and that the amount of energy transmitted from the shock to
the shell also increases.  Unless the shell mass substantially exceeds
the ejecta mass, reflected and transmitted shocks are formed when the
SN shock hits the circumstellar shell.  Vikram demonstrated that the
shock-shell interface is hydrodynamically unstable.  The reflected
shock moves towards the center, and may rebound off the center.
Eventually several shocks may be found criss-crossing the remnant,
leading to a highly complicated interior structure, with more than one
hydrodynamically unstable region possible (Dwarkadas 2000).  A rise in
X-ray emission accompanies each shock-shell collision. When applied to
the observations of SN 1987A, the SN-circumstellar shell model, with
appropriate modifications, confirms the prediction of the outgoing
shock colliding with the circumstellar ring in about 2005 (Chevalier
\& Dwarkadas 1995).\\

{\bf Chris Wright} (ADFA) presented work on ISO observations of the SN
remnant RCW 103.  This supernova remnant has been studied extensively
in the past in the near-infrared (NIR) by Oliva et al. (1989, 1990 and
1999) who showed that the remnant blast wave is interacting with the
interstellar medium and producing very bright emission in lines of
[FeII] and H$_2$. The [FeII] emission coincides with the optical,
radio and x-ray emission, but the H$_2$ emission occurs 20-30
arcseconds outside (i.e. in front) of it. This poses a problem in that
standard shock excitation of H$_2$ predicts that the H$_2$ would
reside either behind or coincident with the optical
emission. Extinction arguments cannot be applied since the extinction
to all of the optical, [FeII] and H$_2$ emission is independently
observed to be the same. Further, the H$_2$ spectrum ``looks''
thermal. Therefore, x-rays have been proposed as a possible excitation
mechanism.  Chris presented ISO observations which covered a large
suite of pure rotational and ro-vibrational H$_2$ lines, out to 28
microns, as well as lines of H, [NeII], [OIV] and [FeII] and the x-ray
sensitive molecules H$_3$+ and HeH+. The latter two were not detected,
and their upper limits may imply interesting constraints on the amount
of x-ray heating. Many H$_2$ lines were detected, and the spectrum
still appears to be shock (i.e. thermally) excited, although more
modelling is required to determine the type of shock. However, there
are several cases where the line appears to have a non-thermal
component to it.\\

{\bf Amy Mioduszewski} (SRCfTA) discussed simulating Radio Images from
Numerical Hydrodynamic Models (Mioduszewski et al.~1997).  She
motivated her discussion by emphasising that while hydrodynamic
simulations are widely used to understand objects such as supernovae
or jets, the calculated pressure, density, and velocity must be linked
to what is observed, the synchrotron radiation from the
material. Assuming minimum energy, Amy demonstrated that the
synchrotron emissivity and opacity can be related to the
hydrodynamical pressure and the number density of the particles.
Using these, she calculated the total synchrotron flux and created an
"image" of the source.  Amy also pointed out that in case of
relativistic jets it is important to consider light travel time
effects, because they significantly influence the appearance of the
jets. In addition she showed that the simulated total intensity light
curves, even of non-evolving jets, are not easily related to the
relatively simple and regular shock structure in the underlying
flow.\\

{\bf John Patterson} (U Adelaide) discussed the potential for using
very high energy gamma rays to understand the high energy
astrophysical processes which occur in objects such as supernova
remnants, gamma ray pulsars and AGN (BL Lacs), as well as the many
unidentified EGRET ($\sim$1 GeV) sources. See Ong (1998) for a review
of the field.  As a leading member of the joint Australian-Japanese
CANGAROO Project at Woomera, John is pushing the frontier of this
ground-based observational area of photons with energies around 100
GeV. These high energy photons are produced in a variety of places by
relativistic processes such as inverse Compton effect and shock
acceleration. A new 10 m Cangaroo II telescope has been commissioned,
and John warmly welcomes co-operation with other Australian facilities
and universities.  \\

\section{Pulsars and the Interstellar Medium I}

The first session on Thursday afternoon opened with {\bf Simon
Johnston} (SRCfTA) reviewing pulsar wind nebulae (PWN).  Typically 1\%
of the spin-down luminosity of pulsars appears as pulsed emission, the
remaining energy presumably coming off in the form of a relativistic
particle wind, which is eventually stopped and shocked by the pressure
of the surrounding interstellar medium (ISM), producing nonthermal
radio emission.  If the space velocity of the pulsar is low, the wind
produces a bubble, or plerion, which can be imaged in radio/optical or
X rays.  On the other hand, if the pulsar has a high space velocity
(and many do -- see the next talk), the rapid motion through the ISM
produces a bow shock, which can be seen in $H_{\alpha}$ or the radio
continuum.  Clearly what will be seen in individual cases will vary
depending on the properties of the pulsar, its space motion, and the
nature of the surrounding ISM. A search for PWN associated with 35
pulsars at 8.4 GHz with the VLA turned up 14 examples.  There appear
to be two classes of pulsars - young systems with
$L_{\mathrm{radio}}/\dot{E} \sim 10^{-4}$ and middle-aged pulsars
where this ratio is below $10^{-6}$.  Several hypotheses could account
for this difference - perhaps the ratio of Poynting flux to particle
flux decreases, or the energy spectrum steepens, or more pulse energy
appears as gamma-rays.  Of course, more observations are needed: a
survey of 50 pulsars with ATCA is underway at 1.4 GHz using pulsar
gating to increase the sensitivity 200-fold.  The survey to date has
turned up one bow shock in 5 pulsars examined -- the Speedboat Nebula
associated with PSR 0906-49.\\

{\bf Matthew Bailes} (Swinburne) gave a talk on the distribution of
pulsar velocities. The first part promoted the new supercomputer
centre at Swinburne, extolling the processing power of the planned
network of 64 linked workstations.  This is impressive, but will be
limited to problems that can be broken into many fairly-independent
parallel modules. The second part discussed pulsar velocities, a topic
that is important in understanding the Galaxy's pulsar population as a
whole. He briefly reviewed the methods for measuring them.  Generally
the old millisecond pulsars have $v\leq 300$ km/s - they are likely
bound to the Galaxy as one would expect.  However, younger pulsars
overall have a higher velocities, which indicates that their
progenitor supernova explosions have a substantial asymmetry.\\

{\bf Lewis Ball} (SRCfTA) spoke about inverse Compton scattering by
relativistic pulsar winds.  Electrons in the wind can upscatter
ambient photons -- starlight or the cosmic microwave background -- to
TeV energies for the expected Lorentz factors $\sim 10^6$.  Generally
this effect is small except for pulsars embedded in strong radiation
fields, such as those in close binary systems.  The pulsar B1259-63,
which is in an eccentric orbit about a Be star, is of particular
interest: conversion of 0.1\% of the pulsar's spin-down luminosity
into 100 GeV photons would give a flux detectable by the CANGAROO II
Cerenkov telescope.  The scattering is a strong function of geometry
and distance from the star -- the pulsar must be rather close to the
star for the effect to be significant.  The distance of B1259-63 to
the Be star ranges from 20 to 300 $R_*$, so the gamma ray luminosity
at periastron is large, with predictable variations around the
well-determined orbit (Kirk, Ball \& Skj\ae raasen 1999; Ball \& Kirk
1999).  Thus CANGAROO II observations will potentially be able to
probe the properties of the pulsar wind, which is otherwise difficult
to detect.\\

{\bf Kinwah Wu} (SRCfTA) presented observations of optical and
infrared lines in the spectrum of the X-ray binary Cir X-1.  The
emission lines are asymmetric, with a narrow component at +350 km/s
and a broader blue-shifted component.  Previously it had been
suggested that the narrow component arises from rotation of an
accretion disc, the corresponding blue-shifted component being absent
because of a shadowing effect at that particular orbital phase.
However, the new observations and archival data show that the profiles
have varied systematically over the last 20 years: the narrow
component always lies in the range 200-400 km/s, while the blue
component varies somewhat both in shape and redshift (Johnston, Fender
\& Wu 1999). Kinwah offered a new model in which the narrow component
is interpreted as arising from the heated surface of the $3-5 {\rm
M}_{\odot}$ companion star, and the broad component arises in an
optically thick outflow driven by super-Eddington accretion from the
neutron star.  The variability in the blue component reflects the
eccentricity of the orbit: at periastron, the companion overfills its
Roche lobe and dumps matter onto the star, producing the outflow.  The
overflow shuts off after periastron; near apastron the remaining
overflow material settles into a quasi-steady accretion disc.  This
model explains the variability of the blueshifted component of the
spectrum and the X-ray behaviour.  One implication of this picture is
that the system has a radial velocity of +430 km/s, which makes Cir
X-1 one of the fastest binaries known.  Even so, a sufficiently
asymmetric supernova explosion can impart the required kick without
unbinding the system (Tauris et al.~1999).\\

\section{Pulsars and the ISM II}

Extreme Scattering Events (ESEs) from pulsars were the topic of {\bf
Mark Walker's} (SRCfTA) presentation. ESEs were first discovered in
extra-galactic sources, the symptoms being a rapid change in flux
density of the observed source. These flux density changes are
attributed to ionised gas clouds in our own Galaxy. From the
observational data, Mark Walker and Mark Wardle have determined the
parameters of these clouds: they have a size of roughly 2 AU, an
electron density of $\sim 10^{3}$ cm$^{-3}$ and a filling factor of
about $5\times 10^{-3}$. They postulate that these clouds may solve
the `missing mass' problem, at least in our Galaxy (Walker \& Wardle
1998; 1999).\\

If a pulsar undergoes an ESE, one can in principle measure three
different quantities. These are the deflection of the image (which
can be measured by VLBI techniques), the delay of the signal (which can
be obtained from pulsar timing) and the magnification of the image.
Pulsars are exceedingly small, and this implies both a large peak
magnification and a large coherent path length. Pulsars are also bright
at low frequencies where the effects should be strongest.
Previous work on ESEs on pulsars include the time delay and flux changes
in the millisecond pulsar PSR B1937+21 and the fringe patterns in the
dynamic spectrum of PSR B1237+25. However, there has been no
systematic observational program carried out and this is needed as a matter
of some urgency.\\

The nature of pulsars means that more information can be gleaned from
ESEs than from say quasars. This in turn will lead to a better
understanding of the structures in the interstellar medium which cause
ESEs.\\

{\bf Jean-Pierre Macquart} (USydney) continued the theme of
scintillations with his presentation on scintillation and density
fluctuations in the ISM.  In scintillation theory it is thought that
energy is deposited at very large scales (kpc or more), that it then
`cascades' down to lower scales before finally dissipating at some
small scale.  However, although this sounds good, the questions of
what provides the energy, how exactly it cascades down and what the
dissipation mechanism is are all unanswered! (see, for example,
Cordes, Weisberg \& Boriakoff 1985) \\

If supernovae are providing the energy at the large scales then
perhaps one might expect to see more turbulence in the vicinity of
supernova remnants. Also, one might expect the power-law index of the
turbulence, $\beta$, to be $\sim$4 rather than the canonical
(Kolmogorov) value of $11/3$.  Is there any observational evidence for
this? In or near the Vela supernova remnant there is some evidence for
$\beta = 4$.  Two surveys of extra-galactic point sources located
behind supernova remnants have been ambiguous with no clear evidence
for a higher power law index although one group do claim an
enhancement behind the Cygnus Loop (Dennison et al.~1984). In summary,
although supernova explosions are the popular choice for the energy
input there is no unambiguous evidence for this (Spangler et
al. 1986). \\

{\bf Jianke Li} (ANU) gave his talk on the topic of the spin-up
mechanism for millisecond pulsars (MSPs). It is widely believed that
MSPs are formed from low-mass X-ray binaries in which a neutron star
accretes matter from its low mass companion. Along with the mass
transfer, the neutron star `accretes' angular momentum causing it to
spin up. Typically, to end up with a 1 millisecond rotation rate
requires the accretion of $10^{-10} M_{\odot}$/yr over $10^{7}$
years. \\

Li argued that even a low magnetic field (say $10^{4}$ Tesla) is
enough to truncate the inner edge of the accretion disk and thus one
has to have a magnetic boundary layer. This magnetic boundary may
impede angular momentum accretion on to the star, so that the angular
momentum accretion could be far less efficient as compared to the
standard model. This casts doubt on whether a low-mass X-ray binary
system such as J1808--369, with a binary period of only two hours, is
indeed spun up by accretion.  \\

\section{Exotics I}
Much of this session was devoted to a discussion of Gamma-Ray bursts
(GRBs) and their relationship to supernovae.  {\bf Ron Ekers} (ATNF)
set the pace with an overview of GRB 980425 and its relationship to SN
1998bw. The 20s outburst was detected by BATSE and localised using
Beppo Sax. Information was quickly distributed on the GRB Coordinates
Network (GCN). Like most well-localised bursts, follow-up radio
observations were carried out using the VLA/ATCA telescopes. About
30\% of GRBs have been detected in the radio, and about 50\% in the
optical. It is interesting that detection in the radio is always
accompanied by the detection of the optical transient. The detection
of a SN within the BeppoSax error box, which has a chance probability
of one in 10$^4$, led to the suspicion that the GRB and SN were
related. If so the SN is the most luminous radio SN known, and the
light curve is quite peculiar. The $\gamma$-ray luminosity was about
10$^{41}$ J. Ron remarked that there had been a suggestion by
Paczynski a few years ago postulating that GRBs could be the result of
hypernovae, so this was one case where theory might have predicted the
observation.\\

{\bf Ray Stathakis} (AAO) presented the results of a cooperative
spectral monitoring campaign of SN 1998bw, carried out on the AAT,
UKST and MSSSO 2.3m, between 11 and 106 days after the Gamma-Ray Burst
(Stathakis et al. 1999). The spectra showed no H, He or Si lines, thus
making it a Type Ic SN. They consisted of broad emission and
absorption features which slowly evolved over the period. SN~1998bw
had entered the supernebular phase by day 106 with the appearance of
nebular emission lines. In comparison to a typical Ic supernova,
SN~1994ai, SN~1998bw was much bluer, and the features were broader and
more distinct at early times. However, transitions and spectral
evolution seen appeared similar, confirming SN~1998bw as a peculiar
type Ic supernova.  While the broader lines (~45\% broader than
classical supernovae at similar epochs) explain much of the
peculiarities of the spectra of SN~1998bw, there is some indication
that additional contribution from line species such as nitrogen,
carbon or titanium may be needed to reproduce the observations. \\

Following this, the GRB 990123 was reviewed by {\bf Brian Boyle}
(AAO). This GRB was first detected by BATSE, and the burst was of 90s
duration. Its brightness was in the top 0.3\% of all BATSE
sources. Optical observations were carried out within 22 secs of the
burst by the ROTSE telescope. Followup Keck spectra were featureless,
apart from some absorption lines, arising perhaps from a foreground
galaxy at z=1.6. The peak V Mag was 8.6, and the total estimated
energy in $\gamma$ rays was about 3.4 $\times$ 10$^{47}$ J. The
luminosity of the optical transient was about 3.3 $\times$ 10$^{16}$
L$_{\odot}$. The host galaxy appears to be a blue star-forming galaxy,
in common with many any other GRB hosts. Brian also summarised briefly
some of the theory of GRBs and the afterglows. The optical decay can
be approximated by three different power laws, due perhaps to the
reverse shock and the forward shocks. The second break may be a
signature of beaming effects.\\

GX 1+4, a low-mass X-ray pulsar toward the galactic centre, was
observed by {\bf Duncan Galloway} (UTas/SRCfTA) with the Rossi X-ray
Timing Explorer (RXTE) satellite during July 1996, $\approx$10 days
before a short-lived ``torque reversal'' event.  Persistent pulsars
such as GX 1+4 typically exhibit no correlation between luminosity
(and hence mass accretion) and spin-up or spin-down rates, contrary to
predictions of existing models. These sources are often found in
``torque states'', where the spin-up or spin-down rate is almost
constant over time-scales of up to 10 years, with torque reversals
occurring irregularly between states. Often the spin-up and spin-down
torques are similar in magnitude.During the {\it RXTE} observation
significant variations in the mean spectrum and the pulse profile were
observed over time-scales of a few hours. Variations of this type have
not previously been observed on such short time-scales, and it is
suggested that these phenomena may be related to the (as yet unknown)
mechanism causing the torque reversals (Galloway et al.~1999; Giles et
al.~1999). \\

\section{Exotics II}
{\bf Dick Manchester} (ATNF) and {\bf Don Melrose} (SRCfTA) talked on
the observations and theory of Anomalous X-ray Pulsars (AXPs).  AXPs
have periods of 6--12 seconds (cf.\ `normal' pulsar periods range from
0.025 s to several seconds), soft X-ray spectra, and relatively low
X-ray luminosities of $10^{28} - 10^{29}$~W --- significantly
below the Eddington limit $10^{31}$~W.  Their X-ray emission is
relatively steady on time scales longer than the pulse period -- much
more so than for accretion-powered binary sources -- and they exhibit
no evidence that they are binary star systems. \\

The pulse periods of AXPs increase with time (Mereghetti, Israel \&
Stella 1998).  If the associated loss of rotational energy is
attributed solely to magnetic dipole radiation, the inferred surface
field is $B \sim 3.2\times 10^{15} (P \dot{P})^{1/2}$~Tesla, whence
$B\sim3\times 10^{10}$~T for typical AXP parameters: $P\sim 10$~s and
$\dot{P}\sim 10^{-11}$ (corresponding to 3 ms/year).  This is much
stronger than the inferred fields for `normal' ($B\sim 10^8$~T) and
millisecond pulsars ($B\sim 10^5$~T).  The idea that their strong
magnetic field may be the defining characteristic of AXPs has led to
them being referred to as `magnetars' (Thompson \& Duncan 1993). \\

More specifically, a magnetar is a neutron star whose surface field
exceeds the critical field strength $B_{\rm c}=4.4\times 10^9$~T at
which the energy corresponding to the cyclotron frequency $\Omega_{\rm
e}$ equals the electron rest energy $\hbar\Omega_{\rm e} = m_{\rm e}
c^2$.  Electric fields of energy densities exceeding that of the
critical field decay spontaneously via electron-positron pair
creation.  Magnetic fields which exceed $B_{\rm c}$ cannot decay in
this way because of kinematic restrictions --- the process of pair
creation would violate momentum conservation. \\

The strong inferred fields of magnetars may arise in one of two ways.
Usov (1992) has shown that the if the strong magnetic fields
associated with some white dwarf stars are frozen in when they
collapse as Type 1a supernovae, then neutron star fields of $10^7$~T
may result.  Duncan \& Thompson (1992) have shown that dynamo action
could generate the inferred fields. \\

The energy loss rates, $4\pi^2 I/(P\dot{P})$ where $I$ is the moment
of inertia, for normal and millisecond pulsars are much higher than
the observed radiation luminosities, and these objects are thought to
be rotation powered.  In contrast, the spin-down luminosity of a
neutron star with $P\sim 10$~s and $\dot{P}\sim 10^{-11}$ is $4\times
10^{25}$~W, much less than the observed X-ray luminosities of AXPs.
It is therefore thought that AXP emission is not powered by rotation,
but rather by the decay of their strong magnetic fields. \\

Some the eight known AXPs are associated with supernova remnants and
some with Soft Gamma-ray Repeaters (SGRs).  There is some evidence
that the AXPs associated with SGRs have the strongest inferred
magnetic fields. The idea that a strong neutron star magnetic field
suppresses radio emission has recently been placed on a more firm
theoretical foundation by Baring and Harding (1998), invoking
suppression of electron-positron pair formation due to increased
photon splitting. \\

The best known SGR was the source of the 5 March 1979 event which
attained a luminosity of $10^{37}$~J and had a clear 8.1~s
periodicity.  It is believed to be associated with a supernova
remnant, N49, in the Large Magellanic Cloud.  A specific model for
this object involves the release of magnetic energy through fractures
of the neutron star crust (Thompson \& Duncan 1995). \\

In a supercritical magnetic field the cross section for the scattering
of radiation with frequencies well below the gyrofrequency is highly
anisotropic.  In particular, scattering of the extraordinary mode is
strongly suppressed with respect to that of radiation in the ordinary
mode.  The consequences of this effect are subtle: it allows
extraordinary mode emission to escape even from close to the neutron
star, and it clearly affects the interpretation of the Eddington
`limit' for accretion powered sources. \\

The Parkes Multibeam Pulsar Survey (Lyne et al., 1999), which has a
flux sensitivity of 150~$\mu$Jy and is seven times more sensitive than
any previous survey, may double the number of radio pulsars from the
750 known before it began.  It has already discovered 362 new pulsars,
including PSR J1814$-$17 which has a period of around 4~s and a high
$\dot{P}$ which places it in the part of $P-\dot{P}$-space occupied by
AXPs.  The AXP 1904$+$09, which has $P=5.16$~s,
$\dot{P}=1.23\times10^{-10}$, and which is associated with
SGR1900$+$14, has recently been claimed as a radio pulsar (Shitov
1999). \\

AXP/SGR/SNR associations, and the relationship between magnetic field
strength and radio emission, may ultimately shed light on the apparent
deficiency of radio pulsars that are associated with supernova
remnants. \\

The collapse of a star and the resulting supernova explosion that
produces a neutron star depends on neutrinos to revive the shock and
eject the outer layers of the star (Bethe \& Wilson 1985).  Four
neutrino flavours are necessary to explain all known neutrino
anomalies, but only three ordinary neutrinos are allowed.  {\bf Yvonne
Wong} (University of Melbourne) is investigating the possibility that
the fourth flavour may arise through oscillations into `sterile'
neutrinos - which do not participate in weak interactions as ordinary
neutrinos do.  The physics of such oscillations, in matter rather than
in vacuo, has important implications for the understanding of
supernova shocks (Nunokawa et al.~1997).\\

{\bf Roberto Soria} (ANU/SRCfTA) and {\bf Amy Mioduszewski} (SRCfTA)
discussed observations of the sources GRS J1655$-$40 and CI Cam which
have answered some questions and raised others.  Optical spectra of
GRS J1655$-$40 display both broad lines in absorption and emission
($>1000\;{\rm km\,s^{-1}}$), and emission lines which are narrower
than the minimum allowed if they originate in an accretion disk.  This
can be explained if the system is a black hole binary, and the narrow
lines originate in an extended envelope surrounding the disk.  The
nature of the source CI Cam remains a mystery.  It has been classified
as a symbiotic star and as a Herbig B object.  It is a bright
emission-line star which exhibited a single uncomplicated X-ray
brightening on 1 April 1998, detected by RXTE and CGRO/BATSE,
brightening from $\sim 0$ to $\sim 2$~Crabs in less than 1 day and
then slowly decaying.  An associated optical brightening by 2
magnitudes was recorded (Fontera et al.~1998).  A radio flare was
detected with the VLBA on day 1 and then at intervals of a few
days. The images, with a resolution of just a few AU show a
slowly-expanding synchrotron shell, with a speed of just $200\;{\rm
km\,s^{-1}}$, and no evidence of the jet-like collimated outflows seen
in all other soft X-ray transient-related radio transients observed
with sufficient resolution.

\section{Masers associated with SNRs}
An overview of the field was presented by {\bf Anne Green} (USydney).
The first observations revealing the likely association of 1720-MHz OH
masers with SNRs were made 30 years ago, but the field then lay
dormant for many years, since the detailed follow up observations of
high spatial resolution and high sensitivity were beyond the reach of
the available instruments.  Interest in the field was revived 5 years
ago (Frail, Goss \& Slysh 1994), with a three pronged attack: first,
high resolution observations of the stronger masers known from the
pioneering work; second, a general single dish survey to see how
widespread the phenomenon was in the light of the better catalogues of
SNRs now available; third, a review of the theoretical explanation and
implications.  To date, about 75 per cent of the known SNRs have been
searched (Frail et al.~1996; Green et al.~1997).  Overall, a 10 per
cent detection rate has been found, although the remnants containing
masers are not uniformly distributed throughout the Galaxy; the
detection rate is higher for SNRs located closer to the Galactic
Centre, where there is a greatly increased density of molecular gas.
Where high resolution observations have been made (with the VLA or
with the ATCA), they reveal clusters of small diameter maser spots
located predominantly at the periphery of the associated SNR.  Zeeman
splitting is often detectable, implying magnetic fields of typically
10$^{-7} T$ or less.  The masers tend to have only a small spread in
radial velocity, irrespective of their location relative to the SNR
boundary, and the inference has been drawn that they occur at points
tangential to the shock front, and thus their velocity represents the
systemic velocity of the remnant itself (but see later).  If so, this
provides a valuable distance indicator, and prompts the theoretical
question of assessing the physical and chemical conditions needed, and
whether the postulated shock provides them. These theoretical aspects
were expanded on by {\bf Mark Wardle} (SRCfTA).  The basic pumping
scheme was suggested 20 years ago and has required only minor
refinements.  It satisfactorily accounts for the fact that the
1720-MHz transition is seen without any accompanying masers at other
OH transitions, occurring at a density too low for their excitation.
More puzzling is how the required OH abundance, densities and
temperatures arise.  Remarkably good progress has been achieved with a
strong consensus that non-dissociative (C-type) shock waves are a key
factor.  More contentious is the question of whether the soft X-ray
emission from the SNR is also vital to the process.\\

Although the framework of understanding is in place, it is largely
based on the very small number of objects studied in detail.  This is
slowly being remedied with follow up observations of more remnants,
and new results were presented by {\bf David Moffett} (U Tasmania).
He found it difficult in several cases to confirm at high resolution
the preliminary detections made with single dishes.  This may be
because in these cases the emission is of a commonly found diffuse
variety of very low gain maser, which lies in the direction of the SNR
purely by chance.  And even where the maser emission was confirmed,
further puzzles arose.  In the case of SNR 332.0+0.2, the velocity is
large and if it represents the systemic velocity, then the implied
distance is unexpectedly large.  So perhaps the velocities can be
significantly offset from the systemic velocity, a possibility that
would reduce their value as distance indicators.  Furthermore the
location of the maser spots is slightly outside the shell as defined
by the radio non-thermal emission; so how well does the radio shell
delineate the outer shock front?  These puzzles highlight the need to
enlarge our sample of well studied objects, since generalisations
drawn from only a few may be quite misleading.\\

Because the collisional excitation is believed to occur as a result of
the SNR shock impinging on an adjacent or surrounding molecular cloud
(Frail et al.~1996; Lockett et al.~1999), one expects to be able to
explore this putative cloud by other means.  For a few objects, this
investigation has begun, and {\bf Jasmina Lazendic} (SRCfTA) described
her work to extend these investigations to more molecular species in a
larger number of remnants, using mm radio observations.  Studies of
molecular hydrogen using IR observations can also be used in such
studies, and work by {\bf Michael Burton} (UNSW) with Jasmina Lazendic
and others have revealed further unexpected phenomena.  The Galactic
object commonly known as the ``snake'' intersects a likely SNR shell
almost at 90 degrees; at the point of intersection is a 1720-MHz
maser, not in itself unexpected since the required shock conditions
could well be fulfilled here. More surprising is the discovery of a
molecular hydrogen outflow jet, apparently emanating from the
intersection point.  The difficulty of accounting for this perhaps
indicates that this is a chance alignment without significance, and
more study is clearly required.\\

Overall, the session was a lively reminder that this field is now
making rapid and exciting progress after a 20 year dormant period while
we waited for the appropriate investigative tools to be developed.\\


%
%





\section*{Acknowledgements}

The organisers would like to thank all the attendees for their
enthusiastic participation in making the workshop such a great
success. A special thanks goes to Samantha Mackinlay who was
responsible for much of the behind-the-scenes work, to Noella D'Cruz
for designing the meeting web page, and to Noella, Amy Mioduszewski
and Michael Rupen for their help in the organisation. We also take
this opportunity to thank the staff of the University Staff Club, and
especially Mark Flusk, for helping things run as smoothly as they did.


\section*{References}





\reference Ball, L., Kirk, J. G. 1999, Astroparticle Phys., in press, (astro-ph/9908201)
\reference Baring, M. G., \& Harding, A. K.~1998, ApJ, {507}, L55-58.
\reference Bartel, N. et al.~1994, Nature,  368, 610
\reference Bethe, H.~A.,  \& Wilson, J.~R.~1985, ApJ, 295, 14 
\reference Chevalier, R.~A., \& Dwarkadas, V.~V.~1995, ApJ, 452, L45
\reference Cordes, J.M., Weisberg, J.M. \& Borkiakoff, V.~1985, ApJ, 288, 221
\reference Dennison, B., et al.~1984, A\&A, 135, 199
\reference Duncan, R. C., \& Thompson, C.~1992, ApJ {392}, L9-13.
\reference Dwarkadas, V.~V.~2000, in preparation
\reference Filippenko, A .V.~1997, ARA\&A, 35, 309 
\reference Fontera, F., Orlandini, M., Amati, L., Dal Fiume, D., Masetti, N., Orr, A.,Parmar, A.N., Brocato, E., Raimondo, G., Piersimoni, A., Tavani, M., \& 
Remillard, R.A.~1998, A\&A, 339, L69
\reference Frail, D. A., Goss, W. M., \& Slysh, V. I.~1994, ApJ, 424, L111.
\reference Frail, D. A., Goss, W. M., Reynoso, E. M., Giacani, E. B., Green,
A. J. \& Otrupcek, R.~1996, AJ, 111,1651
\reference Franco, J., Tenorio-Tagle, G., Bodenheimer, P., \&  Rozyczka, M.~1991, PASP, 103, 803
\reference Gaensler, B. M., Manchester, R. N., Staveley-Smith, L.,
 Tzioumis, A. K., Reynolds, J. E., \& Kesteven, M. J.~1997, ApJ, 479, 845
\reference Galloway D.K., Giles A.B., Greenhill J.G. \& Storey M.C.~1999, MNRAS, in press
\reference Giles A.B., Galloway D.K., Greenhill J.G., Storey M.C. \& Wilson C.A.~1999, ApJ, in press
\reference Green, A. J., Frail, D. A., Goss, W. M. \& Otrupcek, R.~1997, AJ, 114, 2058
\reference Iyudin, A. F., Schonfelder, V., Bennett, K., et al.~1998, Nature 396, 142I
\reference Johnston, H.~M., Fender, R.~P., \& Wu, K.~1999, MNRAS, 308, 415
\reference Kirk, J. G., Ball, L., Skj\ae raasen, O.~1999, Astroparticle Phys., 10, 31
\reference Lockett, P., Gauthier, E. \& Elitzur, M.~1999, ApJ, 511,235
\reference Lyne, A. G., Camilo, F., Manchester, R. N., Bell, J. F., Kaspi,
V. M., D'Amico, N., McKay, N. P. F., Crawford, F., Morris, D. J., Sheppard,
D. C. \& Stairs, I. H.~1999, MNRAS. In press. 
\reference Mereghetti, S., Israel, G. L. \& Stella, L.~1998, MNRAS {
296}, 689-692
\reference Mioduszewski, A.J., Hughes, P.A. \& Duncan, G.C.~1997, ApJ, 476, 649 
\reference Nunokawa, H.,  Peltoniemi, J.~T., Rossi, A.,  \& Valle, J.~W.~F.~1997, Phys.~Rev~D, 56, 1704 
\reference Oliva, E., Moorwood, A.F.M., Danziger, I.J.~1989, A\&A 214, 307 
\reference Oliva, E., Moorwood, A.F.M., Danziger, I.J.~1990, A\&A 240, 453 
\reference Oliva, E., Moorwood, A.F.M., Drapatz, S., Lutz, D., Sturm, E.~1999, A\&A 343, 943 
\reference Ong, R.A.~1998. Phys. Reports, {305}, (Nos3-4), 93-202
\reference Rupen, M.P. et al.~1998, in {\it Radio Emission from Galactic and 
Extragalactic Compact Sources}, ASP Conference Series v. 144, IAU Colloquium 164, eds. J.A. Zensus, G.B. Taylor, \&\ J.M.  Wrobel, p. 355
\reference Shitov, Y. P.~1999, IAU Circular 7110.
\reference Spangler, S.R., et al.~1986, ApJ, 301, 312
\reference Stathakis, R., et al.~1999, MNRAS, in preparation.
\reference Tauris, T.~M., Fender, R.~P., van den Heuvel, E.~P.~J., Johnston, H.~M., \& Wu, K.~1999, MNRAS, in press
\reference Thompson, C. \& Duncan, R. C.~1993, ApJ, {408}, 194-217
\reference Thompson, C., \& Duncan, R. C.~1995, MNRAS, {275}, 255-300.
\reference Usov, V. V.~1992, Nature, 357, 472-474.
\reference Walker, M., \& Wardle, M.~1998, ApJ, 498, L125
\reference Walker, M., \& Wardle, M.~1999, PASA, 16, 262


\end{document}